# Superconductivity in Polycrystalline Diamond Thin Films

Yoshihiko Takano[1*], Masanori Nagao[1], Tomohiro Takenouchi[2],
Hitoshi Umezawa[3], Isao Sakaguchi[1], Masashi Tachiki[1], Hiroshi Kawarada[2]

[1] National Institute for Materials Science, 1-2-1 Sengen, Tsukuba 305-0047 Japan.
[2] School of Science and Engineering, Waseda University, 3-4-1 Okubo, Shinjuku, Tokyo 169-8555, Japan.
[3] Advanced Industrial Science and Technology, 1-1-1, Umezono, Tsukuba 305-8568, Japan.

**Corresponding Author:**
Name: Yoshihiko Takano
Postal address: 1-2-1, Sengen, Tsukuba 305-0047 Japan
Email: takano.yoshihiko@nims.go.jp

**Keywords:**
Boron-doped diamond, superconductivity, semiconductivity, (111) orientation, metal-insulator transition

**Abstract**
Superconductivity was discovered in heavily boron-doped diamond thin films deposited by the microwave plasma assisted chemical vapor deposition (MPCVD) method. Advantages of the MPCVD deposited diamond are the controllability of boron concentration in a wide range, and a high boron concentration, especially in (111) oriented films, compared to that of the high-pressure high-temperature method.  The superconducting transition temperatures are determined to be 8.7K for Tc onset and 5.0K for zero resistance by transport measurements.   And the upper critical field is estimated to be around 7T.

**Introduction**

Diamond is well known as a fascinating jewel. Moreover it is also remarkable for its outstanding physical properties, i.e. the hardest material, the highest thermal conductivity (22K/cmK) and very high Debey temperature (2200K). The highest thermal conductivity originates from the high Debey temperature. According to the McMillan relation, a material having a high Debey temperature such as diamond is a strong candidate for a high transition temperature superconductor if they have enough carriers.

Pure diamond is an electrical insulator. However, diamond becomes a p-type wide-gap semiconductor with boron doping since boron acts as a charge acceptor (*1-2*). Lightly boron-doped diamond is a promising material for semiconducting electronic applications (*3*) such as high frequency and high power devices (*4*) since it has high a breakdown field (>10MV/cm) and a high carrier mobility μ=1600cm$^2$/V-s(hole).

Heavily boron-doped diamond shows an insulator-to-metal transition when the carrier densities exceed around 3e20 cm$^{-3}$. Metallic diamond has been used for electrodes (*5-6*), however, the physical properties have not been studied carefully, particularly at low temperatures. Therefore, the recent news of superconductivity at 2.3K in heavily boron-doped diamond synthesized by high-pressure high-temperature (HPHT) method was received with considerable surprise (*7*). The sample was heated to around 2500-2800K by a graphite heater under 10GPa pressure; this condition is in the direct conversion region, and a catalyst was not used. Thus the growth of large single crystal is difficult.

For application of diamond in devices, it is required that the sample is made into the form of a thin film. Only Chemical Vapor Deposition (CVD) can provide diamond films (*8-9*). The CVD method is essential to developing diamond devices. Recently, we have succeeded in synthesizing superconducting diamond thin films using the MPCVD method. The onset of the superconducting transition is found to be 8.7K, and zero resistance is obtained at 5.0K (10). The superconducting transition temperature is significantly higher than that reported for HPHT diamond (7).

One of the merits of the CVD method is that boron concentration can be artificially controlled over the wide rage. The CVD diamond film can contain boron at relatively high concentration compared to the HPHT method, and the solubility limit of boron into diamond seems to be wider than that of the HPHT method. In this paper we present the growth and superconducting properties of heavily boron-doped polycrystalline diamond thin films deposited by the MPCVD method.

**Experimental**

The heavily boron-doped polycrystalline diamond thin films were deposited on silicon (001) substrates using the MPCVD method with 50 Torr chamber pressure and 500 W microwave power. The substrate temperature was 800-900°C. Hydrogen, methane and trimethylboron were used for source gases. Methane concentration was 1% in hydrogen. The B/C ratio ranged between 2500 and 5000 ppm. The silicon substrates were pretreated by ultrasonic wave using diamond powder. Films of 2-3 μm thickness were obtained after a 8 hr deposition. The boron concentration of the obtained sample was analyzed by using SIMS (CAMECA, IMS-4F, NanoSIMS 50). Hall effect was measured by the standard four-terminal method using a superconducting magnet. Conventional X-ray diffraction was carried out using Cu-Kα raditation. The temperature dependence of the resistivity was observed between room temperature and 1.7K.

**Results and discussion**

An optical microscope image of the obtained film is shown in Fig. 1. The color of the diamond sample is black, because of high boron concentration. Figure 2 shows the x-ray diffraction pattern from 30 to 130 deg. A sharp peak is detected at 2θ = 43.9 degree corresponding to the (111) diffraction of the diamond structure. The (220), (311) and (400) peaks were not detected clearly in this chart. The x-ray diffraction pattern suggested highly (111) orientation of the film.

The temperature dependence of the resistivities for the samples which have several boron concentrations from 1.76e20 to 3.74e21 $cm^{-3}$ are plotted in Fig. 3. The sample with a boron concentration of 1.76e20 $cm^{-3}$ shows semiconducting properties, and its resistivity increases significantly with decreasing temperature, and it becomes insulating at low temperatures. On the other hand, the resistivity of the sample with B=4.25e20 $cm^{-3}$ is significantly suppressed, and the onset of superconducting transition is detected. The metal-insulator transition lies between these two samples, and its boron concentration is found to be around 3e20 $cm^{-3}$. With increasing boron concentration beyond this concentration, the metallic component becomes dominant and the superconducting transition temperature clearly increases. In this plot, the lowest resistivity and highest transition temperature was achieved in the sample with a boron concentration of B=3.75e21 $cm^{-3}$. The carrier concentration of this sample was estimated to be 3.44e21 $cm^{-3}$ by the Hall effect. The boron concentration and carrier concentration were almost the same in this sample.

The temperature dependence of the resistivity of the sample with B=3.75e21 cm$^{-3}$ measured from 10 to 1.7K under several values of magnetic fields up to 9T is shown in Fig. 4. The resistivity is almost independent of temperature above Tc. The resistivity began to drop at around 8.7K (Tc onset) corresponding to the grain where the highest Tc is achieved in this sample. Zero resistivity was obtained at 5.0K (T$_C$ offset) in the zero magnetic field. The superconducting transition temperature is suppressed by the applied magnetic field. The field dependence of the T$_C$ onsets and offsets are plotted in Fig. 5. The extrapolation of the T$_C$ onset gives the value of 10.12 T. Assuming that the superconductivity in diamond is the dirty limit, the upper critical field H$_{C2}$ is estimated to be 7 T, which is similar to the H$_{C2}^{//c}$ of c-axis MgB$_2$ (*11,12*). From the upper critical field of 7T, the coherent length can be estimated to be ξ=6.9nm using the equation of ξ$^2$=Φ$_0$/2πHc$_2$. The irreversibility field is found to be 5.89 T.

We were plotted the superconducting transition temperatures as a function of the carrier density. The carrier densities are determined by the Hall measurements. The superconductivity appeared above the metal-insulator transition around a carrier density of 3e20 cm$^{-3}$, and Tc increases monotonously with increasing carrier densities. The samples we obtained in this work are in the under doped region. Increased carrier densities may provide a higher transition temperature, and an optimum condition may exist at higher carrier densities (13-17).

In conclusion, we have successfully obtained superconductivity in hole-doped diamond with the superconducting transition temperature well above liquid helium temperature in (111) oriented polycrystalline thin films. In order to improve the superconducting properties in diamond, fabrication of single crystal samples is required, particularly homoepitaxial (111) films (18, 19). Diamond can show different electrical properties from insulator, semiconductor, metal and superconductor according to the boron concentration. The developments of new superconducting devices combined with wide-gap semiconducting diamond are anticipated.

. This study was partially supported by Grants-in-Aid from the Japan Society for the Promotion of Science, Ishikawa-Carbon Science and Technology Foundation and Iketani Science and Technology Foundation.

# Figures

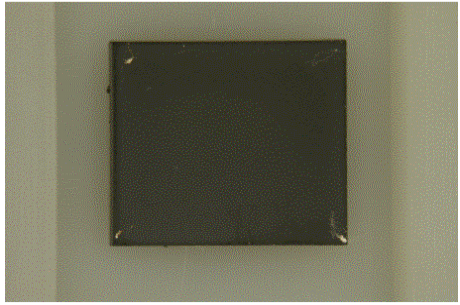

Fig. 1. Optical microscopy image of the polycrystalline diamond thin film deposited by MPCVD method.

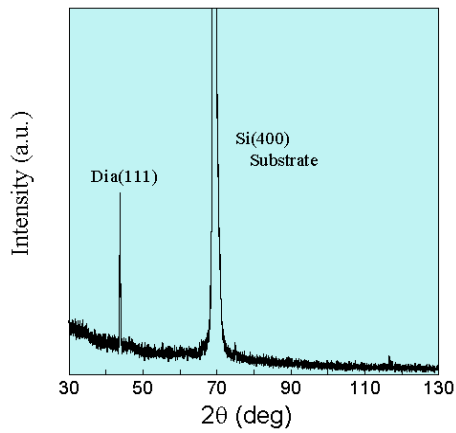

Fig. 2. The X-ray diffraction pattern using Cu-Kα radiation of the polycrystalline diamond film.

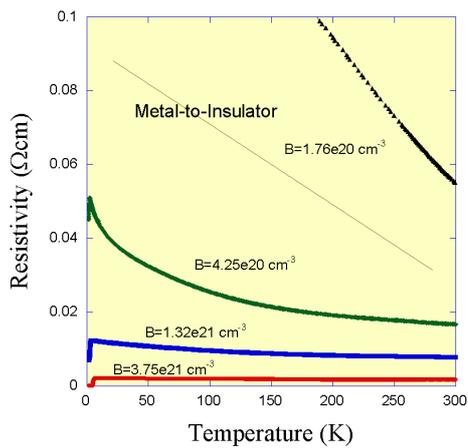

Fig. 3. The temperature dependence of resistivities for the samples having several boron concentrations from 1.76e20 to 3.74e21 cm$^{-3}$.

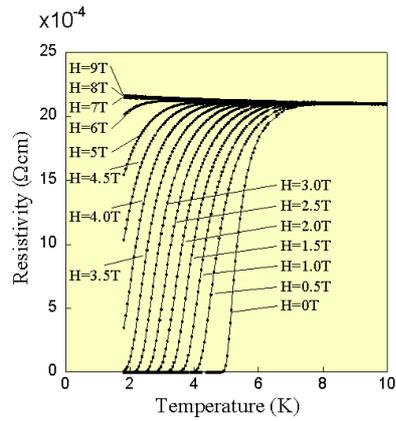

Fig4. Temperature dependence of resistivity under several values of magnetic fields up to 9T. The resistivity began to drop at around 8.7K ($T_C$ onset) and dropped to zero at around 5.0K ($T_C$ offset) in the absence of the field.

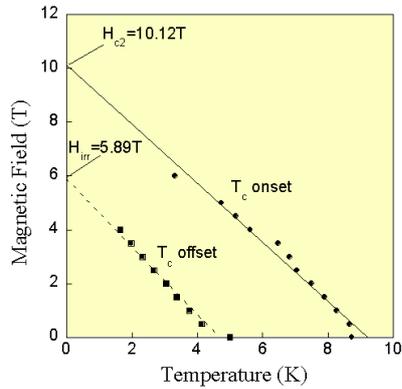

Fig5. The magnetic field dependence of the Tc onsets and offsets. The upper critical field $H_{C2}$ and irreversibility field Hirr are estimated to be 7 and 5.89 T, respectively.

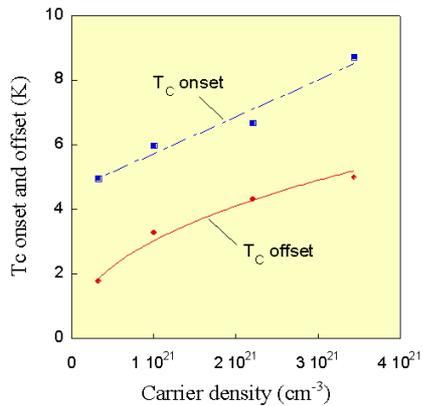

Fig6. The superconducting transition temperatures as a function of the carrier density determined by Hall measurements. The superconductivity appeared above metal-insulator transition around 3e20cm$^{-3}$, and Tc increases monotonously with increasing of carrier densities.